\documentclass[namedreferences]{solarphysics}

\usepackage[hyperref,optionalrh]{spr-sola-addons} 
\usepackage{graphicx}        
\usepackage{color}           

\chardef\us=`\_

\begin{document}

\begin{article}
\begin{opening}

\title{Magnetic Configuration of Active Regions Associated with GLE Events}

\author[addressref={aff1},corref,email={bictr97@gmail.com}]{\inits{R.}\fnm{Regina A.}~\lnm{Suleymanova}}
\author[addressref=aff2,email={leonty@izmiran.ru}]{\inits{L.}\fnm{Leonty I.}~\lnm{Miroshnichenko}}
\author[addressref=aff1,email={vabramenko@gmail.com}]{\inits{V.}\fnm{Valentina I.}~\lnm{Abramenko}}

\address[id=aff1]{Crimean Astrophysical Observatory of Russian Academy of Sciences, Nauchny 298409,
	Bakhchisaray, Republic of Crimea}
\address[id=aff2]{Pushkov Institute of Terrestrial Magnetism, Ionosphere and Radio Wave Propagation
	Russian Academy of Sciences, Moscow 142191, Russia}

\runningauthor{R.~A.~Suleymanova et al.}
\runningtitle{\textit{Solar Physics} Active Regions Associated with GLE Events}

\begin{abstract}
Charged particles, generated in solar flares, sometimes can get extremely high energy, above the 500 MeV level, and produce abrupt ground level enhancements (GLEs) on the ground-based detectors of cosmic rays. The initial flares are strong eruptions and they should be originated from active regions (ARs). A list of GLE events and associated flares was initially available, and our aim here was to find the hosting AR for each GLE event. Moreover, we aimed to classify the revealed ARs using the magneto-morphological classification (MMC: \citealp{2021MNRAS.507.3698A}). We have shown that in 94\% of cases such ARs belong to the most complex morphological classes, namely, $\beta \gamma$, $\beta \delta$, $\gamma \delta$, $\beta \gamma \delta$ classes by the Hale classification and B2, B3 classes by the MMC.
We also found that the GLE-associated ARs are the ARs with the total unsigned magnetic flux much stronger than the common ARs of the same complexity. The set of GLE-related ARs only partially overlaps with the set of SARs (super-active regions). 
These ARs seem to be a manifestation of nonlinearities in the regular process of the global mean-field dynamo, the key ingredient to keep  fluctuations and to create critical conditions in different aspects of the solar activity. 
 
\end{abstract}
\end{opening}
\section{Introduction}
     \label{S-Introduction} 
    Solar active regions (ARs)  are the brightest manifestation of the solar activity. According to the Babcock-Leighton dynamo model, ARs represent flux tubes of the toroidal magnetic field that appear on the solar surface, predominantly  as bipolar structures \citep{1961ApJ...133..572B, 1969ApJ...156....1L}. After reconnection between magnetic flux tubes in the solar atmosphere, an ejection of charged particles occurs producing a solar flare and/or a coronal mass ejection (CME).  Charged particles can penetrate into the outer space as solar cosmic rays (SCR). If accelerated particles have energies in a range from a few dozen keV to few hundred MeV, then these particles are referred to as Solar Energetic Particles (SEPs: \citealp{1946PhRv...70..771F, 1979RvGSP..17.1021D}), and the event of ejection is called as Solar Particle Event (SPE: \citealp{1979RvGSP..17.1021D,1974Ap&SS..28..375S}). Sometimes during a solar flare or CME the energy of SCR exceeds the 500 MeV level. Such events, when invaded in the Earth’s atmosphere, destroy nitrogen and oxygen nuclei, and generate nuclear-cascade process. As a result, we observe an abrupt increase of SCR’s intensity  - the so called Ground Level Enhancement (GLE) event \citep{1946PhRv...70..771F,1979RvGSP..17.1021D}.
     
     The first GLE event was registered on 28 February 1942 \citep{1942TeMAE..47..331L}: ground-based detectors registered the arrival of accelerated protons to the Earth from the Sun. Later, the assumption was made that these events are associated with solar flares \citep{1946PhRv...70..771F}.
     
     Although GLE events are rather rare as compared to the CME occurrence (for example, during Solar Cycle 23 cycle, 12,900 CMEs \citep{2009P&SS...57...53M} and only 16 GLE events were observed  \citep{2011ASTRA...7..439A}), they can contribute significantly into exploring the Solar-Terrestrial relations. In particular, they can give understanding about maximum capabilities of the solar accelerator. These events might be caused by extreme solar flares or CMEs, which, in turn, are related to ARs. During the last decade the interest on GLE events has increased. In particular, it is worth to mention the works by \citet{2022AdSpR..70.2585K}, \citet{2022SoPh..297...75Z}, and \citet{2023SpWea..2103334W}, which explore the relationship between the GLE event and other factors met during the travel to the Earth (such as CMEs, SEP, Heliospheric current sheet, etc.). Properties of flares associated with GLE events were analyzed by \citet{2019ApJ...883...91F}. At the same time, the photospheric source for the GLE event production – the ARs hosting the flare -  also deserves attention. In the present study, we tried to fill this gap.
     
     The aim of this work is to identify and to analyze the ARs that launched the GLE events. We focus on the magneto-morphological properties and the total magnetic flux of these ARs.

\section{Data and Methods}

A method for registration of GLE events is based on ground-based observations of the secondary components of cosmic rays, that are formed as a result of a nuclear-cascade process in the Earth’s atmosphere (neutrons and muons). Ionization cameras, muon telescopes, neutron monitors (NMs) and other detectors can be used for registration. There is a global network that includes over 50 NMs which allow to continuously register the cosmic rays. All data about GLE events are collected in the international Neutron Monitor Data Base (NMBD, \url{www.nmdb.eu/}, \citealp{2011AdSpR..47.2210M}) and featured on the official international website \url{gle.oulu.fi}.

In the present work, the information on the GLE events and flare onset data since 1942 was augmented with information about the hosting ARs and their magneto-morphological and Hale class. The results are gathered in Tables~\ref{Table1} and \ref{Table1.1}.
The identification number for a given GLE event is presented in the first column. 
The second column shows the number of an AR associated with the GLE event. Before 1973, the Royal Greenwich Observatory (RGO) AR number is used according to the Greenwich Photoheliographic Results (GPR) sunspot catalogue \citep{2016SoPh..291.3081B} and is available at \\ \url{www.ukssdc.ac.uk}. For later events, the AR numbers by the United States Air Force/National Oceanic and Atmospheric Administration (NOAA/USAF) Solar Region Summary (SRS) (\url{solarcyclescience.com/activeregions.html}) are presented. In the third column, the Hale class \citep{1919ApJ....49..153H} of an AR is presented for the data since 1982. The Hale class was assigned by SRS of USAF/NOAA (\url{solarcyclescience.com/activeregions.html}).

The magneto-morphological classification (MMC) of an AR attributed to the AR in the present study is shown in the 4th column. The MMC-class was assigned basing on the MMC of ARs suggested recently in \citet{2018Ge&Ae..58.1159A} and  \citet{2021MNRAS.507.3698A}. Details are described below.

The columns from 5th to 7th represent the date of the flare onset, the flare coordinates, the UT time of the optical flare onset, respectively. The optical and X-ray classes (available) are shown in the 8th column. 
 
 We used data of NMBD database for GLE events since 1998 till 2021. For GLE events  since 1970 till 1998, we used information from Solar Geophysical Data (SGD, \url{www.ngdc.noaa.gov/stp/solar/sgd.html}). We also used data from SPE Catalogue for GLE events since 1955 till 1969 \citep{1975cspe.book.....D}. Evaluations published by \citep{1979RvGSP..17.1021D} were used for the earliest GLE events observed in 1942-1949. 

Note, that for some cases, the 5th column contains two dates. These are the cases when the MMC was impossible at the day of the flare onset due to the very close to the limb location of the AR at the moment of the flare. In these cases, the lower date corresponds to the closest day when the center of the AR was closer than 60 degrees to the disk center and classification was performed. 
For the GLE events 16th and 17th we failed to identify any AR which could be responsible for the GLE event.

\section{Identification of an AR Which Could Launch a Flare Followed by a GLE Event }

As it follows from the data in Column 8 of Tables~\ref{Table1} and \ref{Table1.1}, the GLE-associated flares are rather strong ones. As a rule, such flares are launched by ARs, only in very rare cases such flares can occur in a spotless area \citep{1994SoPh..151..169A}. This motivated us to search for ARs which are the most plausible candidates to launch a given flare based on the information about the flare onset and coordinates.     
To identify ARs, we used the following data sources:
\begin{itemize}
	\item Line-of-sight full-disk magnetograms acquired by the Michelson Doppler Imager (MDI) instrument on board the Solar and Heliospheric Observatory (SOHO) telescope, using the Ni I 6768 \AA\ spectral line with the spatial resolution of 4 arcsec (SOHO/MDI/fd: \citealp{1995SoPh..162..129S}).
	\item Space-weather HMI Active Region Patches (sharp\_cea\_720s), namely, line-of-sight magnetograms and continuum images \citep{2014SoPh..289.3549B}, obtained with the Helioseismic and Magnetic Imager (HMI) instrument onboard the Solar Dynamic Observatory (SDO), using the Fe I 6173.3 \AA\ spectral line with the spatial resolution of 1 arcsec (SDO/HMI/SHARP\_CEA: \citealp{2012SoPh..275..229S}, \citealp{2012SoPh..279..295L}).
	\item Drawings and magnetograms, presented on the website of Debrecen observatory (Debrecen Photoheliographic Data, DPD, \url{fenyi.solarobs.csfk.mta.hu/DPD/}: \citealp{2016SoPh..291.3081B}). 
	\item Drawings and sunspot magnetic-field data archive of the Crimean astrophysical observatory of Russian Academy of Sciences \\ (CrAO RAS, \url{sun.crao.ru/observations/sunspots-magnetic-field})
\end{itemize}

To connect a flare with a corresponding AR, we scrutinized a full-disk image (magnetogram and/or sunspot drawing) obtained on the flare onset day. We assumed that the coordinates of the flare and the flare-hosting AR should be very close. In the majority of cases, some AR was observed close to the place of the flare onset. 

In the cases when the flare occurred very close or behind the limb, we went back (or forward) in time assuming that the AR shifts by approximately 13.5 degrees per day. In each case, we found a strong AR appearing on the disk which could launch the considered flare. The appeared AR was assigned to the flare under study. The MMC of such AR was performed for the day when the center of the AR was located closer than 60 degrees to the disk center and the morphological and magnetic structure was well distinguishable. For such cases, the time of classification is shown in the 5th column of Tables~\ref{Table1} and \ref{Table1.1}, beneath the flare time. The time lag between the flare onset and the classification moment was usually 1-2 days and did not exceed 7 days (for GLE39). 

\begin{figure}
	\centerline{\includegraphics[width=1.0\textwidth,clip=]{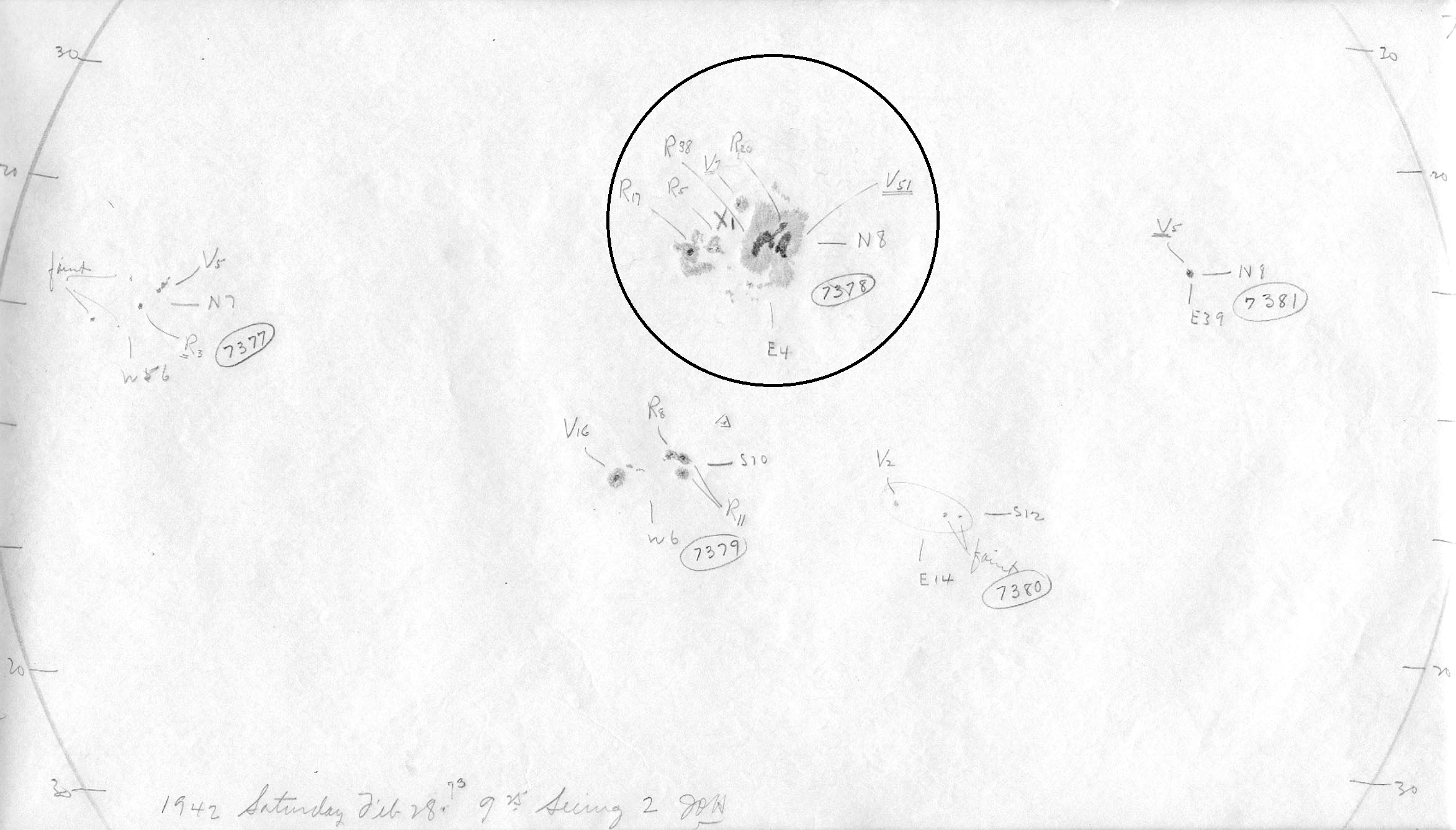}
	}
	\caption{The solar disk drawing acquired on 28 February 1942  used for identification of the GLE01 event. Western limb is on the left. Data provided by Debrecen observatory \citep{2016SoPh..291.3081B}. The AR 14015 is outlined by a circle.}
	\label{fig1}
\end{figure}

\begin{figure}
	\centerline{\includegraphics[width=1.0\textwidth,clip=]{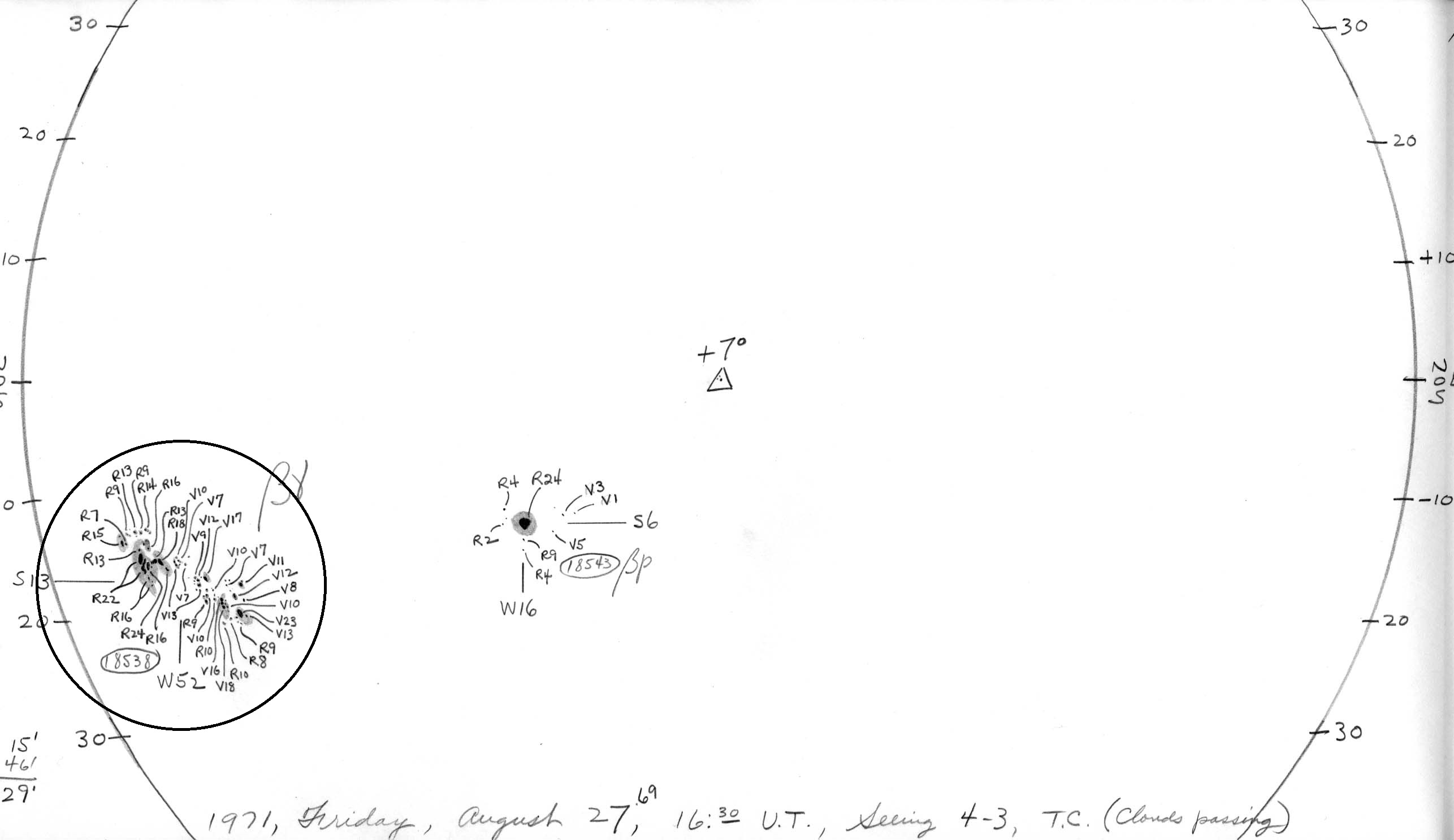}
	}
	\caption{The solar disk drawing acquired on 27 August 1971. Western limb is on the left. Data provided by Debrecen observatory \citep{2016SoPh..291.3081B}. The AR 22877 is outlined by a circle.}
	\label{fig2}
\end{figure}

Examples of AR identification are shown in Figures~\ref{fig1} and~\ref{fig2}. In the first line in Table~\ref{Table1} we see that the event GLE01 was registered on 28 February 1942  an the associated flare has coordinates   07N, 04E. The DPD data base offered the full-disk drawing of sunspots, where on the same coordinates an AR (a strong complex delta-structure) is visible. The corresponding GPR number 14015 is included into the second column of Table~\ref{Table1}. This is the most straightforward identification case in our study. 

 Figure~\ref{fig2} illustrates the identification for one of the most vague cases in our study, namely, the GLE23 event. A flare occurred on 1 September 1971 behind the limb at 11S,120W. A hosting AR was not visible, however 5 days earlier, on 27 August 1971  at the coordinates 13S, 52W  we found the huge bipolar AR 22877, which we attributed to the GLE23. 
 
This way we identified 71 out of 73 GLE events. 

\section{Classification of the Identified ARs.}

For each AR in Tables~\ref{Table1} and \ref{Table1.1} since 1982 the Hale class was attributed  on the basis of the SRS provided by USAF/NOAA. A majority of ARs are complex configurations containing a $\delta$-structure.

Besides the Hale classification, we used in this study the MMC suggested in \citet{2018Ge&Ae..58.1159A},  \citet{2021MNRAS.507.3698A}, \citet{2023MNRAS.518.4746A}.
The underlying idea for the MMC was as follows. Solar dynamo generates the toroidal magnetic field that appears on the surface as bipolar magnetic regions (ARs). A majority of them possess a regular bipolar magneto-morphological structure when an AR follows certain empirical laws, namely, the Hale polarity law, the Joy’s law, the rule of prevalence of the leading spot (\citealp{1961ApJ...133..572B,2015LRSP...12....1V}); regular ARs, class A, in the MMC. However, turbulence, which is unavoidable in the convection zone, can influence the rising toroidal flux tubes. As a result, a variety of distorted structures appear on the surface. A degree of distortion can vary from slight rotation and/or inclination of a bipolar flux tube (irregular ARs of B1 class) to fragmentation and/or strong twisting of a single flux tube (irregular ARs of B2 class), and then to strong chaotic intertwining of several flux tubes (irregular ARs of B3 class).    Unipolar sunspots we segregated into class U. Class A has to subclasses: A1 – ARs that comply with the empirical laws and have no delta-structures, and A2 – the same but with a small delta-structure inside. The presence of small delta-structures turned to be significant in a sense of flaring activity, see \citet{2021MNRAS.507.3698A}.

 According to MMC, strong delta-structures belong to the B2 class. So, the AR 14015 (Figure~\ref{fig1}) associated with GLE01 was classified as B2. Figure~\ref{fig2} shows our single AR of A1-class. Note, however, that in this case, the classification refers to 27 August 1971  whereas the flare that produced GLE23, occurred on   1 September 1971. During 5 days the magnetic structure could change drastically, say, a new delta-structure could emerge. So, the appearence of the most regular A1-class AR in the list of GLE-related ARs is  the most uncertain case in our study. 
 
 We revealed only two ARs classified as B1, namely, the AR 19838/GLE08 which flared on  4 May 1960 at W90 and was classified 3 days earlier. Here, again, the B1 could change to some higher class. And our second B1-case is AR NOAA 8210/GLE56 which possessed the largest sunspot in the following part, so, the leader-prevalence rule was violated and the AR was classified as B1. This AR has beta-delta Hale class, and a moderate size delta-structure between the leading and following parts makes the case a marginal between B1 and B2.   
 
 \section{Results of the Active Regions Analysis }
 
 As mentioned above, 71 (out of 73) GLE events were associated with ARs. The Hale class data were available only for 37 of them (see Table~\ref{Table2}). Only two of them were attributed to the simplest class $\beta$,  the rest  belongs to classes with complex configuration, predominantly with $\delta$-structures.

 The overall outcome of the MMC is shown in Table~\ref{Table3}. Out of total 71 ARs, only one (see Figure~\ref{fig1}) belongs to the A1-class of regular bipolar magnetic structures. The classification was performed 5 days before the flare, so the magnetic structure could become more complicated by the flare onset. There was one unipolar AR (GLE33, 21 August 1979). The rest 69 ARs belong to the complex, irregular classes (B-classes). Also, it is worth mentioning that two $\beta$ Hale class ARs are irregular ARs by MMC. The first one, GLE40 (25 July 1989) belongs to the B3 class because of a complex magnetic structure due to the new magnetic flux. The second one, GLE46 (15 November 1989) was classified as B2 because of a large $\delta$-structure.  The presence of a large $\delta$-structure on 15 November 1989 is obvious on the Sayan Mountains Observatory drawing extracted from the CrAO archive and on the Kislivodsk Observatory filtergram acquired via the DPD site. Moreover, according to USAF NOAA SRS, the next day this AR was classified as BD ($\beta \delta$).
 
 It is also worth noting that, out of the 71 ARs, for 35 ones the class was determined most accurately because the classification was performed for the day of the flare onset.
 A composition of these 35 ARs is: 16 of them are the B2-class-ARs, 17 are the B3-class-ARs, 1 unipolar sunspot and 1 AR of B1-class. So, the majority of them belong to the most complex B2 and B3 classes. 
 
 We also calculated total unsigned magnetic fluxes of analyzed ARs which were observed in 1997 and later (Table~\ref{Table4}). The calculations were carried out by the following way.
 For dataset 1997-2011 the SOHO/MDI/fd magnetograms were used, and SDO/HMI/SHARP\_CEA  magnetograms were taken for the 2012-2021 dataset. The considered ARs were cropped to embrace the AR only, with minimum inclusion of quiet-sun areas. The total unsigned magnetic flux was calculated from the radial component of the magnetic field vector obtained from the line-of-sight component by dividing it over the cosine of the heliocentric angle $\mu$ (\citealp{2001ApJ...555..448H}; more details can be found in \citealp{2023MNRAS.518.4746A}).
 
 We compared the obtained flux data with those obtained earlier in \citet{2023MNRAS.518.4746A} for A1-class ARs, and for an ensemble of (B2+B3)-class ARs, which were observed in Cycles 23 and 24 and are listed in the MMC catalogue  (\url{sun.crao.ru/databases/catalog-mmc-ars}). The results for the average, minimum and maximum values of magnetic fluxes are gathered in Table~\ref{Table5}. The following inferences can be made.
 
 \begin{itemize}
 	\item The average flux of ARs that caused GLE events is 6 times larger than the average flux of A1-class ARs and 2 times larger than the average flux of B2- and B3-class ARs.
 	\item The maximum flux of ARs that caused GLE events is 1.4 times larger than the maximum flux of A1-class ARs and 1.4 times smaller than the maximum flux of (B2+B3)-class ARs.
 	\item The minimum flux of ARs that caused GLE events is 60 and 48 times larger than the minimum flux of A1-class ARs and (B2+B3)-class ARs, respectively.
 \end{itemize}
 
 So, on average, the GLE-associated ARs are the ARs with the total unsigned magnetic flux much stronger than the common ARs of the same complexity.

 \section{Concluding Remarks}
 
 In this work, we identified and analyzed the ARs that launched the GLE events. These ARs were classified by the Hale classification and by the MMC. Statistical analysis showed that in 94\% of cases such ARs belong to the most complex morphological classes, namely, $\beta \gamma$, $\beta \delta$, $\gamma \delta$, $\beta \gamma \delta$ classes by the Hale classification and B2, B3 classes by the MMC. 
 
 We also completed an analysis of the unsigned total magnetic flux of ARs, that launched the GLE events. It was found that the average flux of such ARs is 6 times larger than the average flux of common regular ARs of A1-class and 2 times larger than the average flux of common irregular ARs of the combined set of B2 and B3 classes.
 
 So, one might conclude that the GLE events, in general, are produced by the most complex ARs with the noticeably high magnetic flux. 
 
 During the last decades, the large and complex ARs deserved a special attention is several aspects. For example, in the works by \citet{1988ApJ...328..860B} and \citet{2011A&A...534A..47C} the term super-active region (SAR) was suggested for very strong and complex ARs with extremely high flaring activity. It is believed that such ARs are capable of causing very violent events, therefore, they can have a strong impact on the near-Earth environment. \citet{2011A&A...534A..47C} compiled a list of SARs for the interval containing Solar Cycles 19-23 (1978 - 2006). Out of 45 SARs, 15 ARs turned to be included in the compiled in this work table of GLE events (some ARs prodeced several GLE events). On the other hand, according to our data list, 28 GLE-related ARs were observed during this time interval. So, SARs and GLE-related ARs only partially overlap each other. The reason might lie in unknown underlying physics governing the formation  of SARs and GLE events (for example, the circumstances of the particles propagation in the heliosphere).  
 
 Recently, the notion of ``Rogue" ARs appeared in the literature \citep{2017SoPh..292..167N,2023ApJ...953...51P} to denote ARs that violate the empirical laws compatible with the mean-field dynamo action (the Hale polarity law, the Joy's law) and have a complex and convoluted magnetic configuration (the criteria perfectly coincide with that specified for the irregular ARs of B-class in the MMC). \citet{2017SoPh..292..167N} modeled a solar cycle and included synthetically a ``rogue" AR. These authors have shown that such an addition can drastically change the oncoming cycles down to the complete halt of the dynamo. This experiment justifies the interest in such ARs, not only in the pragmatic sense of activity forecast, but also in the sense of better understanding the dynamo.

 In summary, the solar cycle is caused by the global dynamo. Regular ARs, formed by the global dynamo, fulfill the empirical laws. Nevertheless, turbulence of the convection zone can affect the dynamo work, thus generating complex, non-linear structures, such as, SARs, rogue ARs, and irregular ARs of B-class by the MMC. Their number is noticeably less than that of regular ARs, however, they play an important role in solar activity, namely, they determine the production of extreme flares and CMEs, events of abrupt strong enhancements in the cosmic rays intensity near the Earth, and, moreover, as the simulations show,  they can influence the timing and intensity of oncoming cycles.  Therefore, these ARs are a manifestation of nonlinearities in the regular process of the global mean-field dynamo, the key ingredient to keep  fluctuations and to create critical conditions in different aspects of the solar activity. Undoubtedly, such ARs deserve further investigation, especially, by the multi-wavelength tools. 
 
 \section*{Acknowledgments}
The authors are thankful to the anonymous reviewer for productive discussion and very useful comments which helped to improve the article. The authors are thankful to the staff of the Debrecen observatory for providing the data in the open access.
 
 \begin{dataavailability}
SDO/HMI data are available through the SDO/HMI and SDO/AIA Science Groups. SDO is NASA's Living With a Star (LWS) mission. SOHO/MDI data are available through the SOHO/MDI and SOHO/EIT Science Groups. SOHO is an international collaboration between ESA and NASA. The SDO/HMI and SOHO/MDI data were provided by the Joint Science Operation Center (JSOC). 
 \end{dataavailability}
 \section*{Author contributions}
 RAS, VIA wrote the manuscript, RAS prepared figures and tables, LIM prepared the original GLE data, all authors contributed to the analysis and reviewed the manuscript.
 
 \section*{Declarations}

\subsection*{Competing interests}
The authors declare no competing interests.

\begin{table}
	\caption{GLE events and associated active regions, 1942-1982.}
	\label{Table1}
	\begin{tabular}{cccccccc}     
		\hline                   
		GLE event   & AR   & Hale  & MMC  & Observation &Flare   &Onset&Flare      \\
Number&Number& Class & &   Date      &Position&UT   &H$\alpha$/X\\
\hline
1	  &14015 &	-	 &B2	& 28 Feb. 1942	  &07N 04E &1228 &3+	 \\
2	  &14015 &	-	 &B2	&07 Mar. 1942   &07N 90W &N.O. &--/--\\
&      &       &            & 04 Mar. 1942   &        &     &     \\
3	  &14585 &-	     &B2	&25 Jul. 1946	  &22N 15E &1615 &3+	\\
4 	  &16295 &-	     &B2    &19 Nov. 1949   &	03S 72W&1029 &	3+  \\
&      &       &            &17 Nov. 1949    &        &     &      \\
5	  &17351 &	-    &B3	&23 Feb. 1956   &23N 80W & \textless 0334&	3	\\
&      &       &            &20 Feb. 1956   &        &     &      \\
6     &17597 & -	 &B3	&31 Aug. 1956	  &15N 15E & 1226&	3	\\
7	  &19448 &	- 	 &B2	&17 Jul. 1959	  &16N 31W & 2114&	3+	\\
8	  &19838 &	-	 &B1	&04 May 1960   &13N 90W &1000 &	3	  \\
&      &       &            &01 May 1960   &        &     &         \\
9	  &19998	 &-	    &B2	&03 Sep. 1960   &18N 88E &0037 &	2+	  \\
&      &       &            &05 Sep. 1960	  &        &     &         \\
10	  &20075 &	-	 &B2	&12 Nov. 1960 	  &27N 04W &1315 &3+	  \\
11	  &20075 &	-	 &B2	&15 Nov. 1960	  &25N 35W &0207  &3+	  \\
12	  &20075 &	-	 &B2	&20 Nov. 1960   &28N $\sim$112W&2017 &2	  \\
&      &       &            &16 Nov. 1960	  &        &     &         \\
13	  &20260 &	-	 &B3	&18 Jul. 1961   &07S 59W &0920 &	3+    \\
&      &       &            &17 Jul. 1961	  &        &     &         \\
14	  &20260 &	-	 &B3	&20 Jul. 1961   &06S 90W &1553 &3	\\
&      &       &            &17 Jul 1961	  &        &     &         \\
15	  &20858 &	-	 &B2	&07 Jul. 1966	  &35N 48W  &0025 &	2B	  \\
16	  &-	 &-	     &-	    &28 Jan. 1967	  &22N $\sim$150W&\textless 0200&	--/-- \\
17	  &-	 &-	     &-	    &28 Jan. 1967	  &22N $\sim$150W&\textless 0800&	--/--\\
18	  &21758 &	-	 &B2	&29 Sep. 1968	  &17N 51W	&1617 &	2B	  \\
19	  &21802 &	-	 &B2	&18 Nov. 1968   &21N 87W &\textless 1026 &	1B	  \\
&      &       &            &16 Nov. 1968	  &        &     &         \\
20	  &21894 &	-	 &B3	&25 Feb. 1969	  &13N 37W &0900 &	2B/X2\\
21	  &21930 &	-	 &B2	&30 Mar. 1969   &19N 103W&\textless 0332&	1N	  \\
&      &       &            &26 Mar. 1969	  &        &     &         \\
22	  &22678 &	-	 &B3	&24 Jan. 1971	  &18N 49W &2215 &	3B/X5\\
23	  &22877 &	-	 &A1	&01 Sep. 1971   &11S 120W&\textless 1934&	--/--\\
&      &       &            &27 Aug. 1971	  &        &     &         \\
24	  &23179 &	-	 &B2	&04 Aug. 1972	  &14N 08E &0617 &	3B/X4 \\
25	  &23179 &	-	 &B2	&07 Aug. 1972	  &14N 37W &1449 &	3B/X4\\
26	  &23355 &	-	 &B2	&29 Apr. 1973	  &14N 73W &2056 &	2B/X1\\
&      &       &            &28 Apr. 1973	  &        &     &         \\
27	  &700	 &-	     &B3	&30 Apr. 1976	  &08S 46W &2047 &	2B/X2\\
28	  &889	 &-	     &B3	&19 Sep. 1977	  &08N 57W &\textless 0955&	3B/X2\\
29	  &889	 &-	     &B3	&24 Sep. 1977	  &10N 120W&\textless 0552&	--/--\\
&      &       &            &19 Sep. 1977	  &        &     &         \\
30    &939   &-     &B2     &22 Nov. 1977   &24N 40W &0945 & 2B/X1  \\
31	  &1092	 &-	    &B3	    &07 May 1978	  &23N 72W &0327 &	1N/X2\\
&      &       &            &04 May 1978	  &        &     &         \\
32	  &1294	 &-	    &B2	    &23 Sep. 1978	  &35N 50W &0944 &	3B/X1\\
33	  &1925	 &-	    &U1	    &21 Aug. 1979	  &17N 40W &0550 &	2B/C6\\
34	  &3026	 &-	    &B2 	&10 Apr. 1981	  &07N 36W	&1632&	2B/X2.3	\\
35	  &3079	 &-	    &B2	    &10 May 1981   &03N 75W	&0715&	1N/M1	\\
&      &       &            &08 May 1981	  &        &     &             \\
36	  &3390	 &-	    &B3	    &12 Oct. 1981	  &18S 31E    &0615    &2B/X3.1	\\
37	  &3994	 &$\beta \gamma \delta$	&B3	&26 Nov. 1982   & 12S 87W&0230 &	2B/X4  \\
&      &                      &   &24 Nov. 1982	  &        &     &          \\
38	  &4007	 &$\beta \gamma \delta$	&B2	&07 Dec. 1982   &19S 86W &2341 &1B/X2.8	\\
&      &                      &   &05 Dec. 1982	  &        &     &          \\
\hline
\end{tabular}
\end{table}

\begin{table}
	\caption{GLE events and associated active regions, 1984-2021 (continuation of the table of Table~\ref{Table1}).
	}
	\label{Table1.1}
	\begin{tabular}{cccccccc}     
		\hline                   
		GLE event   & AR   & Hale  & MMC  & Observation &Flare   &Onset&Flare      \\
		Number&Number& Class & &   Date      &Position&UT   &H$\alpha$/X\\
		\hline
39	  &4408 &	$\beta \gamma$      &B3	&16 Feb. 1984   &	--S $\sim$130W&\textless 0858&--/--	\\
&      &                      &         &09 Feb. 1984	  &        &     &          \\
40	  &5603	 &$\beta$               &B3	&25 Jul. 1989   &	26N 85W&0839 &2N/X2	   \\
&      &                      &         &23 Jul. 1989	  &        &     &          \\
41	  &5629	& $\gamma\delta$        &B3	&16 Aug. 1989   &	15S 85W&0058 &	2N/12.5\\
&      &                       &        &13 Aug. 1989	  &        &     &          \\
42	 &5698	&$\beta \gamma \delta$  &B2	&29 Sep. 1989   &	24S $\sim$105W &	1141&	1B/X9\\
&      &                       &        &25 Sep. 1989	  &        &     &            \\
43	 &5747	&$\beta\delta$	&B2     	&19 Oct. 1989	&25S 09E 	&1229	&3B/X13	\\
44	 &5747	&$\beta\delta$	&B3	        &22 Oct. 1989	&27S 32W	&1708	&1N/X2.9\\
45	 &5747	&$\beta\delta$	&B3	        &24 Oct. 1989	&29S 57W	&1738	&2N/X5.7\\
46	 &5786	&$\beta$  	&B2	            &15 Nov. 1989	&11N 28W	&0638	&2B/X3.2\\
47	 &6063	&$\beta \gamma \delta$&B3	&21 May 1990	&34N 37W	&2212	&2B/X5.5\\
48	 &6063	&$\beta\delta$	&B3	        &24 May 1990	&36N 76W	&2046	&1B/X9.3\\
49	&6063	&$\beta\delta$	&B3	        &26 May 1990	&$\sim$35N 103W	&2045	&--/--	\\
&      &                       &        &24 May 1990	  &        &     &            \\
50	&6063	&$\beta\delta$	&B3     	&28 May 1990	&$\sim$35N 120W	&\textless 0516	&--/--	\\
&      &                       &        &24 May 1990	  &        &     &            \\
51	&6659	&$\beta \gamma \delta$	&B3	&11 Jun. 1991	&32N 15W	&0105	&2B/X12	\\
52	&6659	&$\beta \gamma \delta$	&B3	&15 Jun. 1991	&36N 70W	&0633	&3B/X12	\\
&      &                       &        &13 Jun. 1991	  &        &     &            \\
53	&7205	&$\beta\delta$	&B2	        &25 Jun. 1992	&09N 69W	&1947	&1B/M1.4	\\
&      &                       &        &24 Jun. 1992	  &        &     &            \\
54 	&7321	&$\beta \gamma \delta$	&B2	&02 Nov. 1992	&$\sim$25S $\sim$100W	&0231	&--/X9	\\
&      &                       &        &29 Oct. 1992	  &        &     &             \\
55 	&8100	&$\beta \gamma \delta$	&B2	&06 Nov. 1997	&18S 63W	&1149	&2B/X9.4 \\
&      &                       &        &04 Nov. 1997	  &        &     &             \\
56	&8210	&$\gamma\delta$    	&B1	    &02 May 1998	&15S 15W	&1334	&3B/X1.1\\
57	&8210 	&$\beta\delta$	&B3	        &06 May 1998	&11S 65W	&0758	&1N/X2.7	\\
&      &                       &        &05 May 1998	  &        &     &            \\
58 	&8307	&$\beta\delta$	&B3	        &24 Aug. 1998	&18N 09E	&2148	&3B/M7.1	\\
59	&9077	&$\beta \gamma \delta$	&B3	&14 Jul. 2000	&22N 07W	&1003    &3B/X5.7	\\
60 	&9415	&$\beta \gamma \delta$	&B3	&15 Apr. 2001	&20S 85W	&1319 &2B/X14.4	\\
&      &                       &        &13 Apr. 2001	  &        &     &            \\
61 	&9415	&$\beta \gamma \delta$	&B3	&18 Apr. 2001 	&23S W117	&0211	&--/--\\
&      &                       &        &13 Apr. 2001	  &        &     &           \\
62	&9684	&$\beta \gamma$	&B3	        &04 Nov. 2001	&06N 18W	&1603	&3B/1.3 \\
63 	&9742	&$\beta \gamma$	&B3	        &26 Dec. 2001 	&08N 54W	&0432	&--/M7.1\\
&      &                       &        &25 Dec. 2001	  &        &     &\\
64 	&10069	&$\beta \gamma \delta$	&B3	&24 Aug. 2002 	&02S 81W	&0049	&--/X3.1\\
&      &                       &        &21 Aug. 2002	  &        &     &      \\
65	&10486	&$\beta \gamma \delta$	&B3	&28 Oct. 2003	&16S 08E	&1100	&4B/X17.2	\\
66	&10486	&$\beta \gamma \delta$	&B3	&29 Oct. 2003	&19S 09W &2037	&--/X10	\\
67	&10486	&$\beta \gamma \delta$	&B3	&02 Nov. 2003 	&18S 59W &1718	&2B/X8.3\\
&      &                       &        &01 Nov. 2003	  &        &     &  \\
68	&10720&	$\beta\delta$	&B2     	&17 Jan. 2005	&15N 25W	&0659	&3B/X3.8	\\
69	&10720	&$\beta \gamma$	&B2	        &20 Jan. 2005	&14N 61W	&0639 	&2B/X7.1\\
&      &                       &        &19 Jan. 2005	  &        &     &        \\
70	&10930	&$\beta \gamma \delta$	&B2	&13 Dec. 2006&06S 23W	&0217 &4B/X3.4\\
71	&11476	&$\beta \gamma \delta$	&B3	&17 May 2012 	&07N 88W		&0125	&?/M5.1	\\
&      &                       &        &14 May 2012	  &        &     &          \\
72	&12673	&$\beta \gamma \delta$	&B3	&10 Sep. 2017 	&S05 88W	&1906	&X8.2	\\
&      &                       &        &07 Sep. 2017	  &        &     &       \\
73  &12887	&$\beta \gamma$	&B3	        &28 Oct. 2021&26S 07W	&1517 &X1.0 \\

	\end{tabular}
\end{table}

\begin{table}
	\caption{Results of the Hale classification.
	}
	\label{Table2}
	\begin{tabular}{cccccc}     
		\hline                   
		Hale class &$\beta$&$\beta \delta$&$\beta \gamma$&$\gamma \delta$& $\beta \gamma \delta$\\ 		
		\hline
		N	  &2 &	10	 &5	&2	  &18\\
		\hline
	\end{tabular}
\end{table}

\begin{table}
	\caption{Results of the MMC.
	}
	\label{Table3}
	\begin{tabular}{cccccc}     
		\hline                   
		MMC  & A1 & U & B1 & B2 & B3\\ 		
		\hline
		N	  &1 &	1	 &2	&30	  &37\\
		\hline
	\end{tabular}
\end{table}

\begin{table}
	\caption{Characteristics of GLE-ARs occurred during the 1997-2021.
	}
	\label{Table4}
	\begin{tabular}{ccccc}     
		\hline                   
		GLE event   & NOAA   & MMC   & Observation &Flux   \\
		Number&Number&    &   Date      &$10^{22}$ Mx\\
		\hline
		55&	8100&	B2&	04 Nov. 1997&	4.28\\
		56&	8210&	B1&	02 May 1998&	3.12\\
		57&	8210&	B3&	05 May 1998&	2.82\\
		59&	9077&	B3&	14 Jul. 2000&	5.75\\
		60&	9415&	B3&	13 Apr. 2001&	3.40\\
		62&	9684&	B3&	04 Nov. 2001&	3.85\\
		63&	9742&	B3&	25 Dec. 2001&	6.93\\
		64&	10069&	B3&	21 Aug. 2002&	7.03\\
		65&	10486&	B3&	28 Oct. 2003&	10.4\\
		66&	10486&	B3&	29 Oct. 2003&	11.1\\
		67&	10486&	B3&	01 Nov. 2003&	10.1\\
		68&	10720&	B2&	17 Jan. 2005&	6.74\\
		69&	10720&	B2&	19 Jan. 2005&	5.28\\
		70&	10930&	B2&	13 Dec. 2006&	3.51\\
		71&	11476&	B3&	14 May 2012&	4.80\\
		72&	12673&	B3&	07 Sep. 2017&	5.43\\		
		73&	12887&	B3&	28 Oct. 2021&	2.40\\	
		\hline
	\end{tabular}
\end{table}

\begin{table}
	\caption{Unsigned magnetic flux values for active regions of A1 and (B2+B3)-class ARs of Cycles 23 and 24 versus the GLE-associated active regions.
	}
	\label{Table5}
	\begin{tabular}{cccc}     
		\hline                   
		MMC   & Average          & Maximum           & Minimum\\
		 & Flux, $10^{22}$ Mx& Flux, $10^{22}$ Mx & Flux, $10^{22}$ Mx\\
		\hline
		A1	  &0.92              &	7.83             &0.04\\
		B2+B3 &2.62& 15.28&0.05\\
		GLE-ARs&5.71&11.14&2.40\\
		\hline
	\end{tabular}
\end{table}

 {}

\end{article} 

\end{document}